\documentclass[a4paper,11pt]{article}
\usepackage{pos}
\usepackage{nicefrac}
\usepackage{comment}

\newcommand{\bs}[1]{\boldsymbol{#1}}

\newcommand*{\rom}[1]{\expandafter\@slowromancap\romannumeral #1@}

\newcommand{\Q}{Q}

\newcommand{\fig}{Fig.~}

\newcommand{\tstar}{T_*}
\newcommand{\der}{\mathrm{d}}

\newcommand{\vb}{\vec}
\renewcommand{\vec}[1]{\mathrm{\mathbf{#1}}}

\newcommand{\tauT}{\tau_{\mathrm{BMSS}}}
\newcommand{\eps}{\varepsilon}
\title{Heavy quark diffusion coefficient during hydrodynamization - non-equilibrium vs. equilibrium}

\author[a]{K.~Boguslavski}
\author[b]{A.~Kurkela}
\author[c,d]{T.~Lappi} 
\author[a]{F.~Lindenbauer} 
\author*[c,d,e]{J.~Peuron} 
\affiliation[a]{Institute for Theoretical Physics, Technische Universit\"{a}t Wien, 1040 Vienna, Austria}
\affiliation[b]{Faculty of Science and Technology, University of Stavanger, 4036 Stavanger, Norway}
\affiliation[c]{Department of Physics, P.O.~Box 35, 40014 University of Jyv\"{a}skyl\"{a}, Finland}
\affiliation[d]{Helsinki Institute of Physics, P.O.~Box 64, 00014 University of Helsinki, Finland}
\affiliation[e]{Dept. of Physics, Lund University,  S\"{o}lvegatan  14A, Lund,SE-223 62, Sweden}
\emailAdd{kirill.boguslavski@tuwien.ac.at}
\emailAdd{aleksi.kurkela@cern.ch}
\emailAdd{tuomas.v.v.lappi@jyu.fi}
\emailAdd{jarkko.t.peuron@jyu.fi}
\emailAdd{florian.lindenbauer@tuwien.ac.at}

\abstract{We compute the heavy quark momentum diffusion coefficient using effective kinetic theory for a
system going through bottom-up isotropization until approximate hydrodynamization. We find that
when comparing the nonthermal diffusion coefficient to the thermal one for the same energy density,
the observed deviations throughout the whole evolution are within 30\% from the thermal value.
For thermal systems matched to other quantities we observe considerably larger deviations. We also observe that
the diffusion coefficient in the transverse direction dominates at large occupation number, whereas
for an underoccupied system the longitudinal diffusion coefficient dominates. Similarly, we study the jet quenching parameter, where we obtain a smooth evolution connecting the large values of the glasma phase with the smaller values in the hydrodynamical regime.}

\FullConference{ HardProbes2023\\
 26-31 March 2023 \\
 Aschaffenburg, Germany\\}

\begin{document}
\maketitle

\section{Introduction}
Recent studies on transport coefficients out of equilibrium have indicated that the glasma stage can have considerable impact on the coefficients \cite{Avramescu:2023qvv,Carrington:2022bnv,Carrington:2021dvw,Carrington:2020sww,Boguslavski:2020tqz,Ipp:2020nfu,Khowal:2021zoo,Sun:2019fud}. However, there has been a literature gap until very recently: equilibrium transport coefficients are relatively well known but the evolution of transport coefficients during hydrodynamization remained poorly understood. We report here of our recent studies where we aimed to close the gap and investigated the heavy quark momentum diffusion coefficient $\kappa$ \cite{Boguslavski:2023fdm} and the jet quenching parameter $\hat{q}$ \cite{Boguslavski:2023alu} during hydrodynamization using effective kinetic theory. 

The two main questions these proceedings address involve the magnitude of $\kappa$ compared to its equilibrium value during hydrodynamization and  the relative importance of transverse and longitudinal diffusion coefficients during the hydrodynamization process.

\section{Method: effective kinetic theory and bottom-up thermalization}
\begin{figure}
\centering
\includegraphics[scale=0.4]{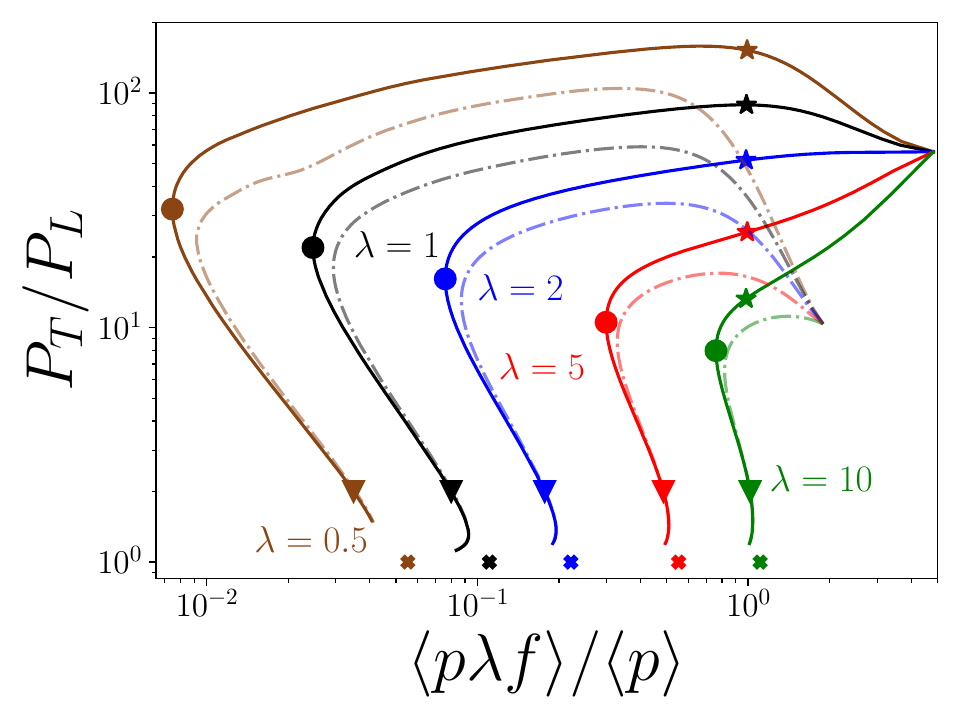}
\caption{Trajectory of the system during the bottom-up thermalization on the occupation number anisotropy plane. Solid and dashed curves correspond to different initial conditions. Reproduced from \cite{Kurkela:2015qoa}.}
\label{fig:kurkelazhu}
\end{figure}

We reproduce the bottom-up thermalization \cite{Baier:2000sb} scenario using effective kinetic theory \cite{Arnold:2002zm}, as in \cite{Kurkela:2015qoa}. The evolution of the system is illustrated in \fig \ref{fig:kurkelazhu}. In order to make a connection to the evolution of the system and other quantities, we have placed a few markers in \fig \ref{fig:kurkelazhu}. The star marker is placed at $f = \nicefrac{1}{\lambda} = \nicefrac{1}{\left(4 \pi N_c \alpha_s \right)},$ where $N_c$ is the number of colors and $\alpha_s$ is the strong coupling constant. For weak couplings this also corresponds to maximum anisotropy. The circle marker is placed at minimum occupancy, which in the bottom-up thermalization picture corresponds to $\alpha_s$.  Finally, the triangle marker is placed at  $\nicefrac{P_T}{P_L} = 2$, corresponding to approximate isotropy. The crosses at the bottom correspond to the expected values at thermal equilibrium. 

In effective kinetic theory the dynamical degree of freedom is the gluon phase space density 
$
f(\bs{p}) = \frac{1}{\nu_g}\frac{\der N}{\der^3 x\, \der^3 \bs{p}},
$
whose time-evolution is given by the Boltzmann equation
\begin{align}
\label{eq:EKTEOMS}
 -\frac{\partial f(\bs{p})}{\partial \tau } = \mathcal{C}_{1 \leftrightarrow 2  }[f(\bs{p})] + \mathcal{C}_{2 \leftrightarrow 2  }[f(\bs{p})] - \frac{p_z}{\tau} \frac{\partial}{\partial p_z} f(\bs{p}).
\end{align}

The dominant contribution to the diffusion coefficient $\kappa$  arises from scattering with the medium gluons via t-channel gluon exchange \cite{Moore:2004tg}. The coefficient is given by
\begin{align}
\label{eq:KT_omGen}
3\kappa= \frac{\left< \Delta k^2 \right>}{\Delta t} =  \frac{1}{2M}\int_{\bs{k} \bs{k^\prime} \bs{p^\prime}}\left(2 \pi \right)^3 \delta^3\left( \bs{p} +\bs{k} - \bs{p^\prime} - \bs{k^\prime} \right)   2 \pi \delta \left(k^\prime - k \right) \bs{q}^2 
\left[ \left| \mathcal{M}_\kappa \right|^2 f(\bs{k}) (1+f(\bs{k^\prime})) \right],
\end{align}
where $k, k^\prime$ are the ingoing and outgoing gluon momenta,  $q = k - k^\prime,$ is the momentum transfer and $p, p^\prime$ are the ingoing and outgoing heavy quark momenta. The integration measure is given by $\int_{\bs{p}} =  \int \nicefrac{\der p^3}{2 p^0\left(2 \pi \right)^3}$. The matrix element corresponding to this process is 
$
 \left|\mathcal{M} \right|^2_{\mathrm{\kappa}} = \left[N_c C_H g^4 \right] \frac{16 M^2 k^2 \left( 1 + \cos^2 \theta_{\bs{k}\bs{k}'} \right)}{(q^2+m_D^2)^2}.
$
The effective temperature of the infrared modes is \\ $
\tstar = \frac{2 \lambda}{m_D}  \int \nicefrac{\der^3 p}{\left(2 \pi \right)^3} f(p) (1+f(p)),$ where the Debye screening mass is $m_D^2 = 4 \int \nicefrac{\der^3 p }{(2 \pi)^3} \nicefrac{\lambda f(p)}{p }$. 
When comparing equilibrium and nonequilibrium systems, we need an estimate for the temperature of the corresponding equilibrium system. This temperature is defined through energy density $\eps$ as $
 T_\eps = \left(\nicefrac{30\,\eps}{\pi^2\nu_g} \right)^{1/4}.$
In equilibrium the quantities above $(\tstar, m_D, T_\eps)$ are computed using the Bose-Einstein distribution.

\section{Results}
Since there is no unambiguous way to compare equilibrium and nonequilibrium systems, we will try to compare the equilibrium and nonequilibrium systems for the same $m_D,$ $\tstar$ and $\eps$. The comparison is done as a function of time. As a consequence,  the corresponding thermal system changes during the time-evolution.  We rescale the time with the thermalization timescale $\tauT =  \nicefrac{\alpha_s^{\nicefrac{-13}{5}}}{\Q_s}$. 

\begin{figure}
\includegraphics[scale=0.3]{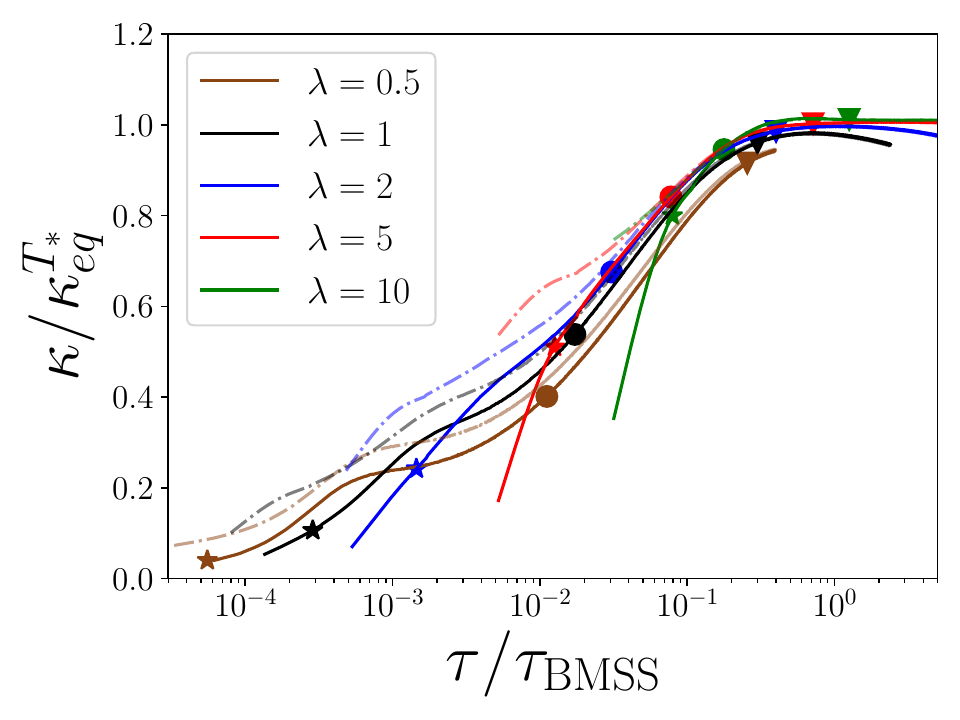}
\includegraphics[scale=0.3]{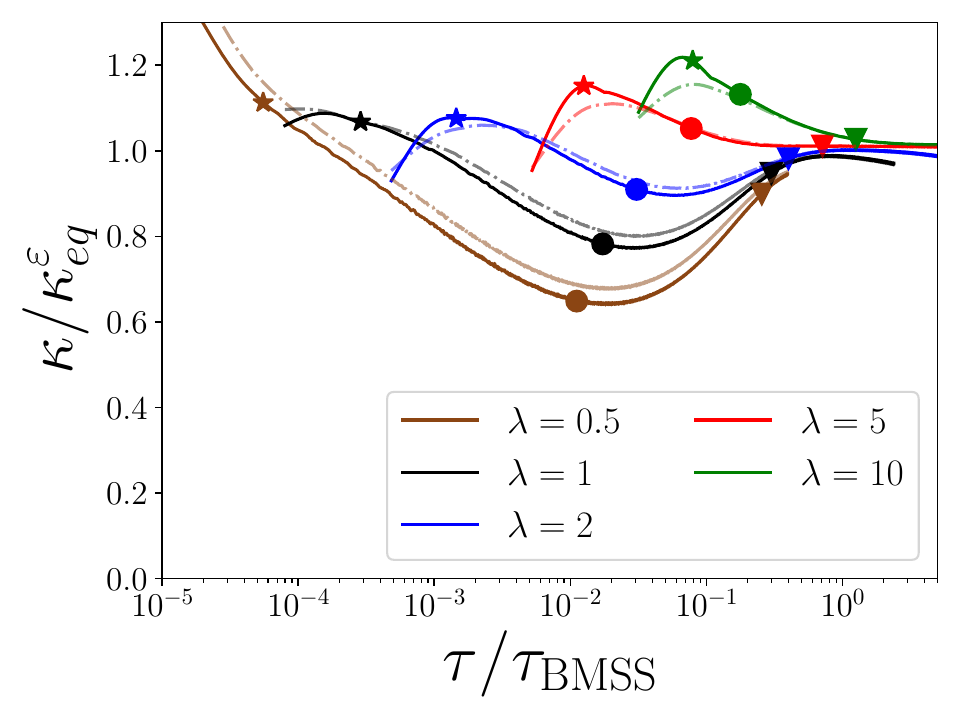}
\includegraphics[scale=0.3]{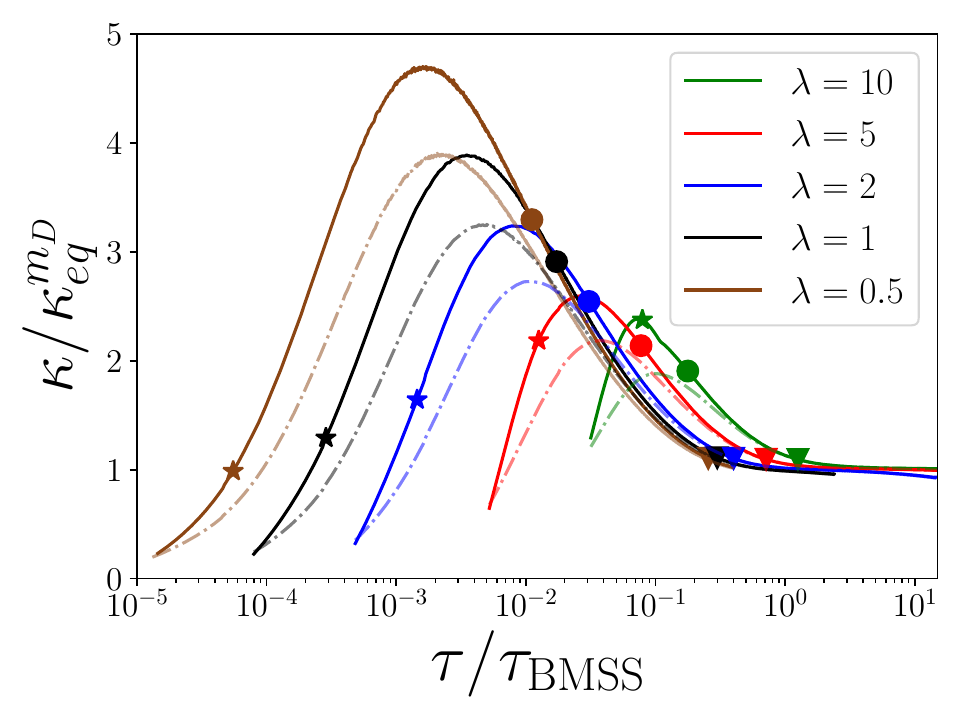}
\caption{Three different ways to compare equilibrium and nonequilibrium. Left: comparing for the same effective temperature of the infrared modes. Center: for the same energy density. Right: for the same screening mass. }
\label{fig:kappacomparison}
\end{figure}

The results are shown in \fig  \ref{fig:kappacomparison}. We observe that when matching for the same screening mass $m_D$ or infrared temperature $\tstar$ there are large deviations during the equilibration. However when matching for the same energy density $\eps$ the deviations are (depending on the coupling) within approximately $\sim 30\%$ during the evolution. Thus matching for the same energy density (Landau matching) is the best way to compare equilibrium and nonequilibrium systems in this case.

\begin{figure}
\centering
\includegraphics[scale=0.3]{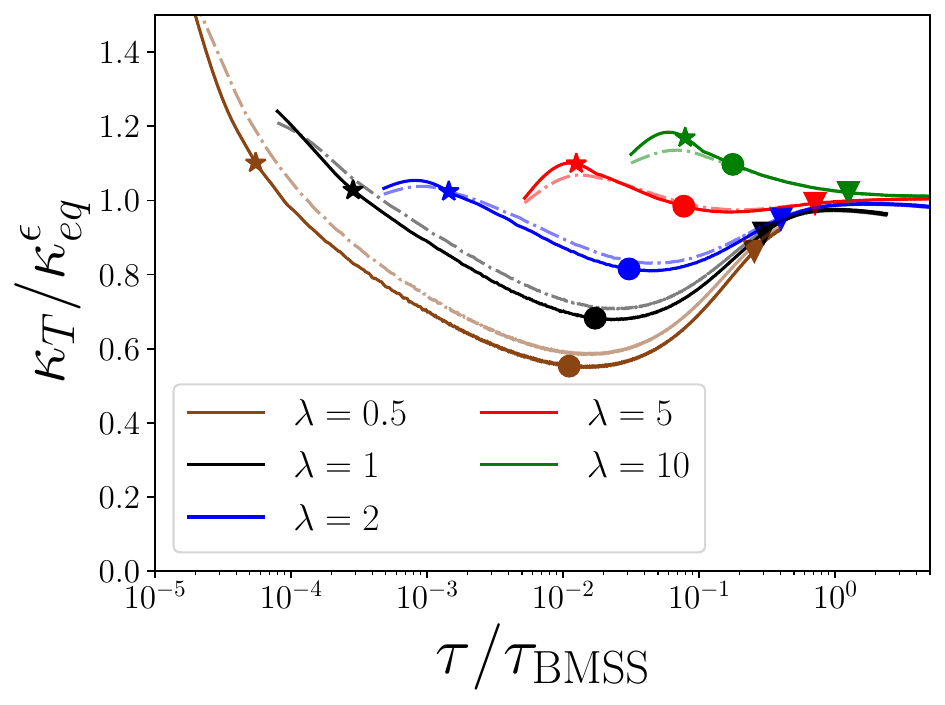}
\includegraphics[scale=0.3]{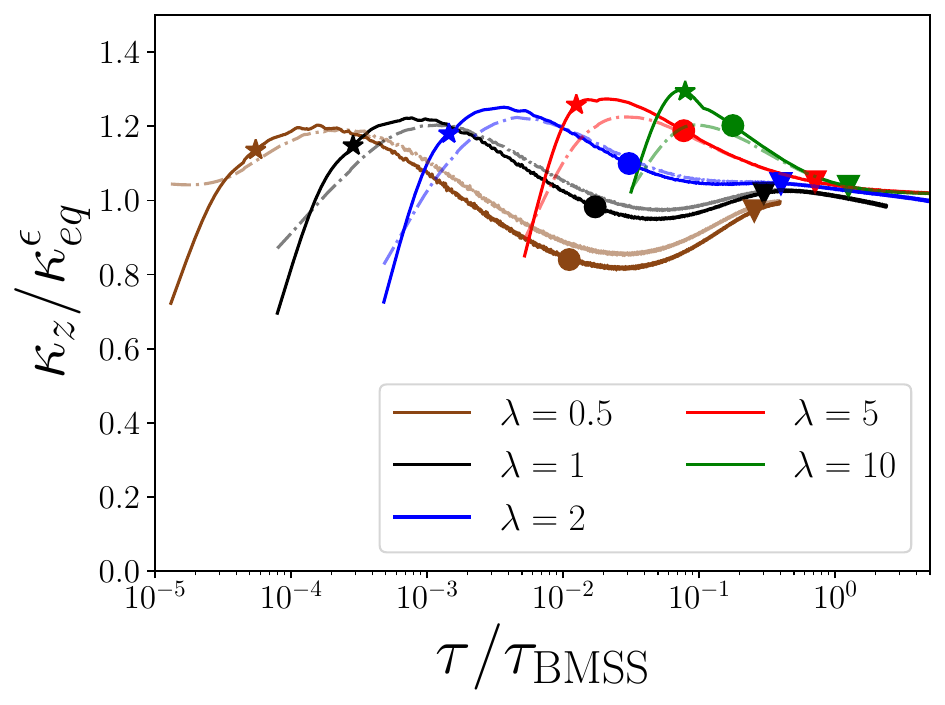}
\caption{Transverse and longitudinal diffusion coefficients compared to their equilibrium values for the same energy density.}
\label{fig:kappatranslong}
\end{figure}

We can also break the comparison down into transverse and longitudinal components as we have done in \fig \ref{fig:kappatranslong}. We observe that the transverse ($\kappa_T$) and longitudinal ($\kappa_L$) diffusion coefficients behave qualitatively similarly to the full coefficient (except in the case of the longitudinal diffusion coefficient at very early times). For smaller coupling $\lambda$ we observe larger deviations. This is most likely due to the fact that for small coupling the bottom-up thermalization is reproduced more accurately. 

\begin{figure}
\centering
\includegraphics[scale=0.35]{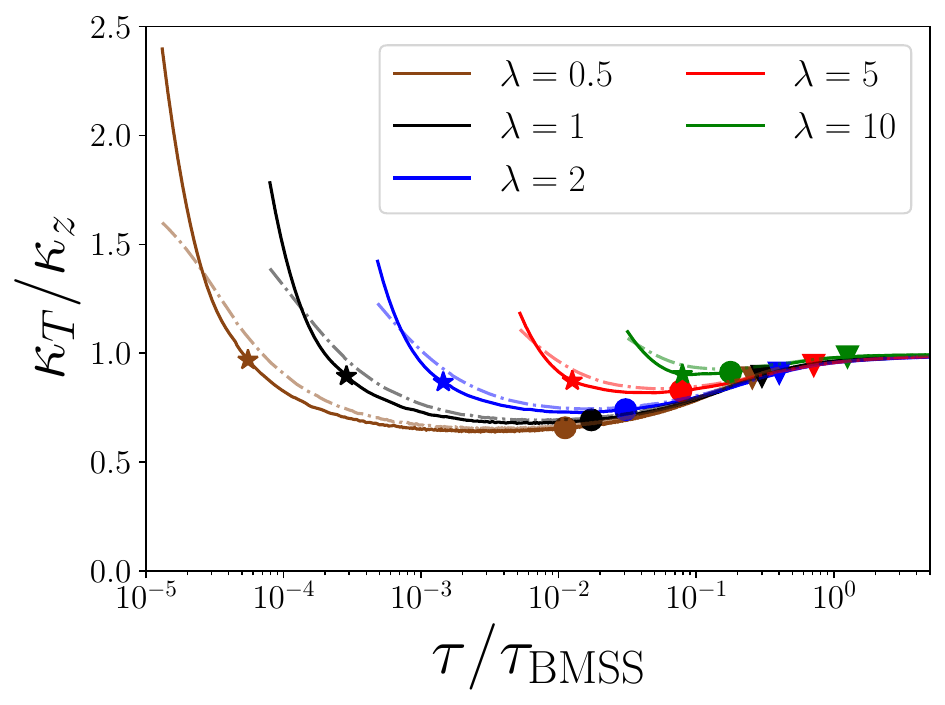}
\caption{Ratio of the transverse and longitudinal diffusion coefficients during the hydrodynamization process. }
\label{fig:transvslong}
\end{figure}

\fig \ref{fig:transvslong} shows the ratio of transverse and longitudinal diffusion coefficients during the evolution. We observe that the transverse diffusion coefficient is initially enhanced compared to the longitudinal coefficient. When the system becomes underoccupied, the hierarchy is inverted, and the longitudinal coefficient is enhanced compared to the transverse coefficient. The anisotropy of the coefficients can become sizable, of the order 10 - 40 \%, depending on the coupling strength. The plot also shows the emergence of a limiting attractor, which we will discuss elsewhere in more detail.

\begin{figure}
\centering
\includegraphics[scale=0.37]{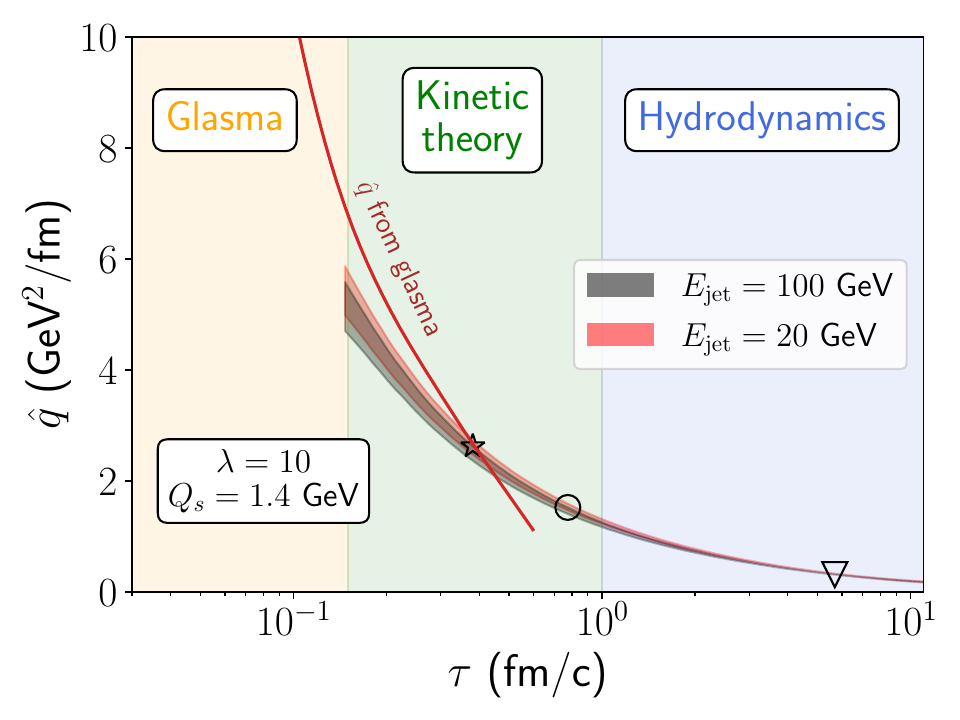}
\caption{The value of the jet quenching factor $\hat{q}$ computed according to the procedure described in the text. }
\label{fig:qhat}
\end{figure}

Then we proceed to the jet quenching factor $\hat{q}$ defined as $    \hat{q}^{ij}=\frac{\der \langle q^i q^j \rangle }{\der L}$. Here we use the following convention:  $\hat{x}$ jet direction, $\hat{z}$ beam direction. The jet quenching factor is given by  
\begin{align}
\hat q^{ij} &= \frac{1}{4d_R}\lim_{|\vb p|\to\infty}\int_{\substack{\vb k\vb k'\vb p'\\q_\perp < \Lambda_\perp}} q_\perp^i q_\perp^j (2\pi)^4\delta^4(P+K-P'-K') \frac{\left|\mathcal M_{ag}^{ag}\right|^2}{|\vb p|} f_{\vb k}\left(1+ f_{\vb k'}\right),
\label{eq:qhat_general}
\end{align}
where $\left|\mathcal M_{ag}^{ag}\right|^2$ is the matrix element corresponding to elastic scatterings off  in-medium gluons. Here we consider a quark jet. However the value of $\hat{q}$ for a gluon jet can be obtained by scaling with a simple Casimir factor. The curves shown in \fig \ref{fig:qhat} are obtained as follows: We match $\varepsilon$ to glasma as in \cite{Ipp:2020nfu} to obtain the value of  $Q_s$ at the initial condition. Then   $\hat{q}$ is matched to the result of  JETSCAPE \cite{JETSCAPE:2021ehl}
 at the triangle marker to obtain a value for the transverse momentum transfer cutoff $\Lambda_\perp$ at that time. The bands correspond to  different cutoff models and initial conditions. We observe that our results match the glasma simulation at early times relatively well and smoothly connect to the hydrodynamic evolution.

\section{Conclusions}
The two main conclusions of these proceedings are, that during the hydrodynamization  $\kappa$ is within 30 \% from its equilibrium value when the equilibrium and nonequilibrium systems are matched for the same energy density. The second conclusion is that there is a clear hierarchy between transverse and longitudinal diffusion coefficients. Initially the transverse diffusion coefficient $\kappa_T$ dominates. At underoccupation $\kappa_z$ is larger. In both cases the deviation is roughly a factor of two. 

Our results may be used for phenomenological descriptions of heavy quark diffusion, quarkonium dynamics and jet quenching. Our future plans involve studying limiting attractors using $\kappa$ and $\hat{q}$ as test observables.

\acknowledgments
The authors would like to thank N.~Brambilla, M.~Escobedo, D.I.~Müller, A.~Rothkopf and M.~Strickland for valuable discussions. 
This work is supported  by the European Research Council, ERC-2018-ADG-835105 YoctoLHC.  This work was also supported under the European Union’s
Horizon 2020 research and innovation  by the STRONG-2020 project (grant agreement No. 824093). The content of this article does not reflect the official opinion of the European Union and responsibility for the information and views expressed therein lies entirely with the authors.  This work was funded in part by the Knut and Alice Wallenberg foundation, contract number 2017.0036. TL and JP have been supported by the Academy of Finland, by the Centre of Excellence in Quark Matter (project 346324) and project 321840.
KB and FL would like to thank the Austrian Science Fund (FWF) for support under project P 34455, and FL is additionally supported by the Doctoral Program W1252-N27 Particles and Interactions.
The authors wish to acknowledge CSC – IT Center for Science, Finland, for computational resources.  We acknowledge grants of computer capacity from the Finnish Grid and Cloud Infrastructure (persistent identifier urn:nbn:fi:research-infras-2016072533 ).
 The authors wish to acknowledge the Vienna Scientific
Cluster (VSC) project 71444 for computational resources.

\bibliographystyle{JHEP-2modlong}
\bibliography{spires}

\providecommand{\href}[2]{#2}\begingroup\raggedright\begin{thebibliography}{10}

\bibitem{Avramescu:2023qvv}
D.~Avramescu, V.~B\u{a}ran, V.~Greco, A.~Ipp, D.~I. M\"uller and M.~Ruggieri,
  {\it {Simulating jets and heavy quarks in the Glasma using the colored
  particle-in-cell method}},  \href{http://arXiv.org/abs/2303.05599}{{\tt
  arXiv:2303.05599 [hep-ph]}}.

\bibitem{Carrington:2022bnv}
M.~E. Carrington, A.~Czajka and S.~Mrowczynski, {\it {Transport of hard probes
  through glasma}},  \href{http://arXiv.org/abs/2202.00357}{{\tt
  arXiv:2202.00357 [nucl-th]}}.

\bibitem{Carrington:2021dvw}
M.~E. Carrington, A.~Czajka and S.~Mrowczynski, {\it {Jet quenching in
  glasma}},  \href{http://dx.doi.org/10.1016/j.physletb.2022.137464}{{\em Phys.
  Lett. B} {\bf 834} (2022) 137464}
  [\href{http://arXiv.org/abs/2112.06812}{{\tt arXiv:2112.06812 [hep-ph]}}].

\bibitem{Carrington:2020sww}
M.~E. Carrington, A.~Czajka and S.~Mrowczynski, {\it {Heavy Quarks Embedded in
  Glasma}},  \href{http://dx.doi.org/10.1016/j.nuclphysa.2020.121914}{{\em
  Nucl. Phys. A} {\bf 1001} (2020) 121914}
  [\href{http://arXiv.org/abs/2001.05074}{{\tt arXiv:2001.05074 [nucl-th]}}].

\bibitem{Boguslavski:2020tqz}
K.~Boguslavski, A.~Kurkela, T.~Lappi and J.~Peuron, {\it {Heavy quark diffusion
  in an overoccupied gluon plasma}},
  \href{http://dx.doi.org/10.1007/JHEP09(2020)077}{{\em JHEP} {\bf 09} (2020)
  077} [\href{http://arXiv.org/abs/2005.02418}{{\tt arXiv:2005.02418
  [hep-ph]}}].

\bibitem{Ipp:2020nfu}
A.~Ipp, D.~I. M\"uller and D.~Schuh, {\it {Jet momentum broadening in the
  pre-equilibrium Glasma}},
  \href{http://dx.doi.org/10.1016/j.physletb.2020.135810}{{\em Phys. Lett. B}
  {\bf 810} (2020) 135810} [\href{http://arXiv.org/abs/2009.14206}{{\tt
  arXiv:2009.14206 [hep-ph]}}].

\bibitem{Khowal:2021zoo}
P.~Khowal, S.~K. Das, L.~Oliva and M.~Ruggieri, {\it {Heavy quarks in the early
  stage of high energy nuclear collisions at RHIC and LHC: Brownian motion
  versus diffusion in the evolving Glasma}},
  \href{http://dx.doi.org/10.1140/epjp/s13360-022-02517-w}{{\em Eur. Phys. J.
  Plus} {\bf 137} (2022)~no.~3 307}
  [\href{http://arXiv.org/abs/2110.14610}{{\tt arXiv:2110.14610 [hep-ph]}}].

\bibitem{Sun:2019fud}
Y.~Sun, G.~Coci, S.~K. Das, S.~Plumari, M.~Ruggieri and V.~Greco, {\it {Impact
  of Glasma on heavy quark observables in nucleus-nucleus collisions at LHC}},
  \href{http://dx.doi.org/10.1016/j.physletb.2019.134933}{{\em Phys. Lett.}
  {\bf B798} (2019) 134933} [\href{http://arXiv.org/abs/1902.06254}{{\tt
  arXiv:1902.06254 [nucl-th]}}].

\bibitem{Boguslavski:2023fdm}
K.~Boguslavski, A.~Kurkela, T.~Lappi, F.~Lindenbauer and J.~Peuron, {\it {Heavy
  quark diffusion coefficient in heavy-ion collisions via kinetic theory}},
  \href{http://arXiv.org/abs/2303.12520}{{\tt arXiv:2303.12520 [hep-ph]}}.

\bibitem{Boguslavski:2023alu}
K.~Boguslavski, A.~Kurkela, T.~Lappi, F.~Lindenbauer and J.~Peuron, {\it {Jet
  momentum broadening during initial stages in heavy-ion collisions}},
  \href{http://arXiv.org/abs/2303.12595}{{\tt arXiv:2303.12595 [hep-ph]}}.

\bibitem{Baier:2000sb}
R.~Baier, A.~H. Mueller, D.~Schiff and D.~T. Son, {\it 'bottom-up'
  thermalization in heavy ion collisions},  {\em Phys. Lett.} {\bf B502} (2001)
  51 [\href{http://arXiv.org/abs/hep-ph/0009237}{{\tt arXiv:hep-ph/0009237}}].

\bibitem{Arnold:2002zm}
P.~B. Arnold, G.~D. Moore and L.~G. Yaffe, {\it Effective kinetic theory for
  high temperature gauge theories},
  \href{http://dx.doi.org/10.1088/1126-6708/2003/01/030}{{\em JHEP} {\bf 01}
  (2003) 030} [\href{http://arXiv.org/abs/hep-ph/0209353}{{\tt
  arXiv:hep-ph/0209353 [hep-ph]}}].

\bibitem{Kurkela:2015qoa}
A.~Kurkela and Y.~Zhu, {\it Isotropization and hydrodynamization in weakly
  coupled heavy-ion collisions},
  \href{http://dx.doi.org/10.1103/PhysRevLett.115.182301}{{\em Phys. Rev.
  Lett.} {\bf 115} (2015) 182301} [\href{http://arXiv.org/abs/1506.06647}{{\tt
  arXiv:1506.06647 [hep-ph]}}].

\bibitem{Moore:2004tg}
G.~D. Moore and D.~Teaney, {\it How much do heavy quarks thermalize in a heavy
  ion collision?},  \href{http://dx.doi.org/10.1103/PhysRevC.71.064904}{{\em
  Phys. Rev.} {\bf C71} (2005) 064904}
  [\href{http://arXiv.org/abs/hep-ph/0412346}{{\tt arXiv:hep-ph/0412346
  [hep-ph]}}].

\bibitem{JETSCAPE:2021ehl}
{\bf JETSCAPE} collaboration, S.~Cao {\em et.~al.}, {\it {Determining the jet
  transport coefficient qhat from inclusive hadron suppression measurements
  using Bayesian parameter estimation}},
  \href{http://dx.doi.org/10.1103/PhysRevC.104.024905}{{\em Phys. Rev. C} {\bf
  104} (2021)~no.~2 024905} [\href{http://arXiv.org/abs/2102.11337}{{\tt
  arXiv:2102.11337 [nucl-th]}}].

\end{thebibliography}\endgroup

\end{document}